\documentclass[twocolumn,pra]{revtex4}
\usepackage{dcolumn}
\usepackage{graphicx}

\providecommand{\U}[1]{\protect\rule{.1in}{.1in}}
\newcommand{\ds}{\displaystyle}

\newcommand{\be}{\begin{equation}}
\newcommand{\ee}{\end{equation}}
\newcommand{\beq}{\begin{eqnarray}}
\newcommand{\eeq}{\end{eqnarray}}

\newcommand{\w}{\omega}
\newcommand{\W}{\Omega}
\newcommand{\g}{\gamma}
\newcommand{\G}{\Gamma}

\newcommand{\ket}{\rangle}
\newcommand{\bra}{\langle}

\newcommand{\bnn}{\begin{eqnarray*}}
\newcommand{\enn}{\end{eqnarray*}}

\newlength{\textwidthm}
\begin{document}

\title{Intensity correlations in resonance nonlinear
magneto-optical rotation }
\author{Tigran S. Varzhapetyan$^{1,2}$, Hebin Li$^1$,
Gombojav O. Ariunbold$^1$, Vladimir A. Sautenkov$^{1,3}$, %D. Sarkisyan$^2$,
Yuri V. Rostovtsev$^1$, and Marlan O. Scully$^{1,4}$}
\affiliation{$^1$Institute for Quantum Studies and Department of Physics, Texas A\&M
University, College Station, Texas 77843, USA}
\affiliation{$^2$Institute for Physical Research of NAS of Armenia, Ashtarak-2
  378410, Armenia}
\affiliation{
   $^3$Lebedev Institute of Physics, Moscow 119991, Russia
}
\affiliation{$^4$Princeton Institute for the Science and Technology of Materials and Department
of Mechanical \& Aerospace Engineering. Princeton University, Princeton, New
Jersey 08544, USA}
\keywords{one two three}
\pacs{PACS number}

\date{\today}

\begin{abstract}

We have studied the intensity correlations between two orthogonally
linearly polarized components of a laser field
propagating through a resonant atomic medium.
These experiments have been performed in a Rubidium atomic vapor.
We observe that
the correlations between the orthogonally polarized components of
the laser beam are maximal in the absence of a magnetic field.
The magnitude of the correlations depends on the applied magnetic field,
and the magnitude
first decreases and then increases with increasing magnetic field.
Minimal correlations
and maximal rotation angles are observed at the same magnetic fields.
The width of the correlation function is directly
proportional to the excited state lifetime and
inversely proportional to the Rabi frequency of laser field.
These results can be useful for improving optical
magnetometers and for optical field or atomic spin squeezing.

\end{abstract}

\maketitle

\section{Introduction}

Quantum coherence effects have been a focus of research activities for the
last two decades, because they may drastically change the optical properties
of a medium. For example, electromagnetically induced transparency (EIT)
\cite{Harris, MarangosRev2005}, predicted and observed in CW and pulsed
regimes~\cite{kochar-eit,vas04pra}, practically allows absorption to vanish.
The nonlinear response of a resonant atomic medium at moderate
optical intensities can be strongly enhanced by creating
maximal coherence between long-lived sublevels of the ground state
\cite{BudkerRev2002, MarangosRev2005, harris-max, vas04pra2, Weis1992}.
The corresponding steep dispersion results in
the ultra-slow or fast propagation of light
pulses~\cite{Hau1999, Kash1999, Budker1999}, which can produce huge optical
delays~\cite{delay} and can be used for drastic modification of
the phase-matching conditions for Brillouin scattering~\cite{sound},
and four-wave mixing~\cite{rost06prl}.
It is possible to manipulate a coherent medium and produce
optical pulses at rates faster
than the relaxation rates of the medium~\cite{switching}, in order
to develop bright sources for
efficient generation of IR and  FIR pulses~\cite{kalugin06ol}.
Also, the nonlinear properties of such media are enhanced, allowing
the implementation of quantum light storage~\cite{Lukin2001, Hau2001},
nonlinear optics at a few photon level~\cite{Harris1998},
and other nonlinear effects~\cite{lukin-hemmer}.

It is worth noting here that media with quantum coherence might
be used for the sensitive measurement of magnetic fields~\cite{scully01prl}.
Recent interest in nonlinear magneto-optical
effects (NMOE), such as nonlinear magneto-optical rotation (NMOR)
\cite{BudkerRev2002} or nonlinear polarization self-rotation (NPSR)
\cite{Budker2001pra}, has been fueled by possible applications
to precision magnetometry
\cite{Budker2000pra, Weis2005ELO, Hollberg2005, VAS2000}.

The amplitude noise of optical fields increases
after the fields interact with an atomic media (atomic excess noise)
\cite{Yabuzaki1991, Walser1993, Boyd1995, Camparo1998, Camparo1999}.
Several processes, such as the conversion of phase noise to amplitude noise
and the four-wave mixing process,
can be associated with the generation of atomic noise
\cite{Yabuzaki1991, Walser1993, Camparo1998, Camparo1999, Boyd1995,
Zibrov2005}.
The strong coupling of optical fields in $\Lambda$-type atoms provides a way
of controlling amplitude fluctuations.
The amplitude correlations have been studied in EIT
\cite{AlzarEPL2003, VAS2005pra, vas07jmo,Ari2006, Lezama2006} and
NMOE \cite{Lezama2004} experiments.
Observation of sub-Poisson statistics
of the amplitude noise
is reported in
\cite{AlzarEPL2003}. Amplitude correlations and anti-correlations
are observed in two orthogonally circularly polarized optical beams from
the same laser \cite{VAS2005pra} and in orthogonally linearly polarized
optical beams from two independent lasers \cite{Lezama2006}.
Power spectra of intensity fluctuations was studied in \cite{vas07jmo}.
Noise spectroscopy of nonlinear magneto-optical resonances in Rb vapor
at a detection frequency 2.5 MHz with bandwidth 30 kHz is performed
in \cite{Lezama2004}, where the photon noise limit has been reached.

Coherent effects may be used to reduce the noise level below the
limit given by photon fluctuations.
Several schemes for squeezing of the fields are
proposed and studied in EIT \cite{Roch1997, Lukin1999, Pinard2006prl} or
NPSR \cite{Matsko2002, Matsko2003, Lvovsky2003, Zibrov2005, Pinard2006pra}
configurations. It has been demonstrated that
the atomic excess noise can reduce
squeezing in the fields~\cite{Pinard2006pra},
and the addition of excess noise to quantum state of radiation
is important and should be taken into account in
experiments involving coherently prepared atoms. For instance,
recently a considerable influence of the excess noise on light storage
has been observed \cite{Hsu2006prl}.

In this paper, we have considered the atomic noise correlations in
an NMOR experiment with a rubidium vapor.
In a previous paper~\cite{VAS2005pra},
we studied the fluctuations between left and right
circularly polarized beams, which are the normal modes of the system.
Here, we have studied the fluctuations between two orthogonally
linearly polarized beams, which are not the normal modes of the system.
Our current experimental configuration is widely used in optical
magnetometry measurements~\cite{BudkerRev2002}, and therefore furnishes an
ideal system for studying these fluctuations. We have observed the variation of
the correlation function $G^{(2)}(\tau)$  ($\tau$ is the delay time) as a
function of the magnitude of an applied magnetic field.

The paper is organized as follows. In Section II, we describe
the experimental setup and the results. In Section III,
we present a theoretical model, which gives us an explanation of
the results. Finally, in Section IV, we present our conclusions.

\section{Experimental setup and obtained results}

The schematic of our experimental setup is shown in Fig.~1a.
An external cavity diode laser ~\cite{vasilliev2006}
(the linewidth is less than 1 MHz)
is tuned to the center of the Doppler broadened $D_1$ line
(transition $5S_{1/2}(F=2)\rightarrow 5P_{1/2}(F=1)$) of
$^{\text{87}}$Rb (see Fig.~1b).
After a polarizing beam splitter (PBS), the laser beam
(of diameter $0.1$~cm and power $0.9$~mW)
enters a glass cell (of length $7.5$~cm)
filled with an atomic Rubidium vapor (with natural isotope abundance and
atomic density $10^{12}$ cm$^{-3}$).

\begin{figure}[tb] %fig2
\center{
\includegraphics[width=.8\columnwidth,angle=270]{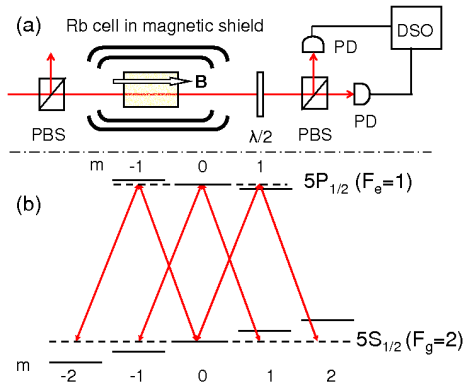}
}
\caption{%Figure 1.
Experimental setup (a) and energy levels (b). The ground state
Zeeman splitting of $Rb^{87}$ is 0.7 MHz/G. }
\end{figure}

The cell is placed inside a two-layer magnetic shield
that protects the cell from uncontrolled environmental magnetic fields.
A solenoid is installed inside the shield
to create a longitudinal magnetic field $B$.
A half-wave plate is placed between
the two PBSs to rotate the polarization of the output beam to 45$^{\text{o}}$
with respect to transmission axis of the second PBS.
This angle is set by making the two beams have equal intensities
at a large one-photon detuning from the atomic resonance in
the cell at room temperature.
Two orthogonally linearly polarized beams from the second PBS are focused on
identical fast  photo-detectors (PDs) with a frequency bandwidth
of 75  kHz-1.2 GHz. The optical path lengths for both beams and length
of cables between the PDs and the data recorder are chosen to be the same to avoid a time delay between the signals
in two registration channels. The signals from PDs are recorded by
a digital storage oscilloscope (DSO) with a bandwidth of 100 MHz.

We have also studied the transmission and
the polarization rotation of the beams as a function of
the two-photon detuning by scanning the magnitude of the longitudinal
magnetic field $B$ at a frequency of several Hz. For this,
the fast photo-detectors have been replaced by low-frequency detectors
to monitor the output from the second beam splitter.
Then, using the recorded signals, $S_1$ and $S_2$, we calculate
the transmission by
\be
T = \left({S_1 + S_2\over S_{01} +S_{02} }\right),
\ee
and the polarization rotation by
\be
\phi = \arcsin\left({S_1 - S_2\over S_1 + S_2}\right),
\ee
where $S_{01}$ and $S_{02}$ are the signals from the photo-detectors
without atomic resonance absorption.

%\section{Obtained results}
\begin{figure}[htb] %fig2
\center{
\includegraphics[width=.9\columnwidth]{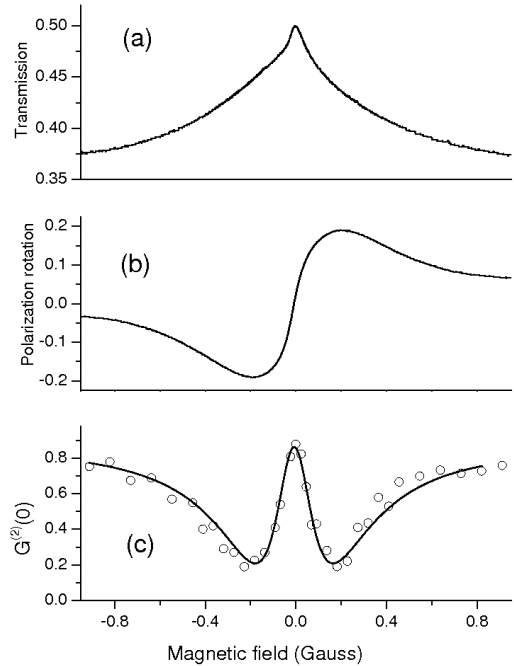}
}
\caption{%Figure 2.
Magnetic field dependence of the EIT (Fig.~2a), polarization
rotation (Fig. 2b) and correlation function $G^{(2)}(0)$) (Fig. 2b).
In Fig.~2(c) the circles represent experimental data and the solid
curve is a visual guide for eyes. }
\end{figure}

We start the presentation of the results with
a demonstration of the narrow EIT resonance.  The narrow EIT resonance
is related to the ``dark state'' of Rb atoms formed by the action
of two laser fields.
The transmission $T$ versus magnetic field $B$ is shown in Fig.~2a.
In a magnetic field, the linear polarization of the laser field
rotates, and the polarization rotation dependence on
the external magnetic field is shown in Fig.~2b.
The width of nonlinear magnetic optical resonances
is determined by the power broadening of the two-photon transition,
which for our conditions is narrower than the natural optical width (6 MHz).
The observed background in Fig.~2b is due to the linear rotation
of polarization.

To study the fluctuations of an optical field transmitted through a dense Rb
vapor, we have registered the time dependent signal fluctuations
$\delta S_{1,2}(t)$ of the two optical beams after the second PBS.
We record these
signal fluctuations for the two output beams in a 10~$\mu s$ time-window
for different magnetic fields, and then we
calculate the normalized correlation functions given by
\be %\[
G^{(2)}(\tau)=\frac{\left\langle \delta S_{1}(t)\delta S_{2}(t+\tau
)\right\rangle }{\sqrt{\left\langle [\delta S_{1}(t)]^{2}\right\rangle
\left\langle [\delta S_{2}(t+\tau)]^{2}\right\rangle }},
\ee %\]
where $\delta S_{1,2}(t)$ are the time dependent fluctuations
of the two  beams, and stochastic averaging~\cite{book}
denoted by angular brackets is defined as
$\langle Q(t) \rangle \equiv 1/T\int^{t+T}_t Q(t') dt'$.

We are unable
to detect any intensity fluctuations when the cell
has been removed from the laser beams.
Of course, it is well-known that diode laser radiation possesses
low intensity noise.
In our experiments, the photo-detector noise dominates,
as in Refs.~\cite{VAS2005pra,vas07jmo}.

In the presence of a resonant medium, however, the situation
changes, and the phase noise of a diode laser
is transformed into intensity fluctuations.
The magnetic dependence of the correlation function at zero time delay,
$G^{(2)}(0)$, is shown in Fig.~2c.
As one can see, a correlation magnitude of close to 0.9
is obtained at zero magnetic field.

Increasing the magnetic field up to 0.2~G results in decreasing
the correlation function magnitude to $\sim 0.2$. But further increase
of the magnetic field leads to a revival of the correlation, and
it reaches $\sim 0.7$ at 0.8~G field strength. We underline here
that this behavior is different from that observed in \cite{VAS2005pra}.
The key difference is that, in this paper, we study the fluctuations of
the orthogonal linear polarization components, which are not normal modes
because of the Faraday effect.

By examining the polarization angle rotation dependence
shown in Fig.~2b, one can conclude that
the maximum correlation is obtained when the magnetic
field and the polarization rotation in the atomic medium are close to zero.
The correlation functions $G^{(2)}(\tau)$ for three different magnetic
fields are shown in Fig.~3.

\begin{figure}[tb] %fig3
\center{
\includegraphics[width=.9\columnwidth]{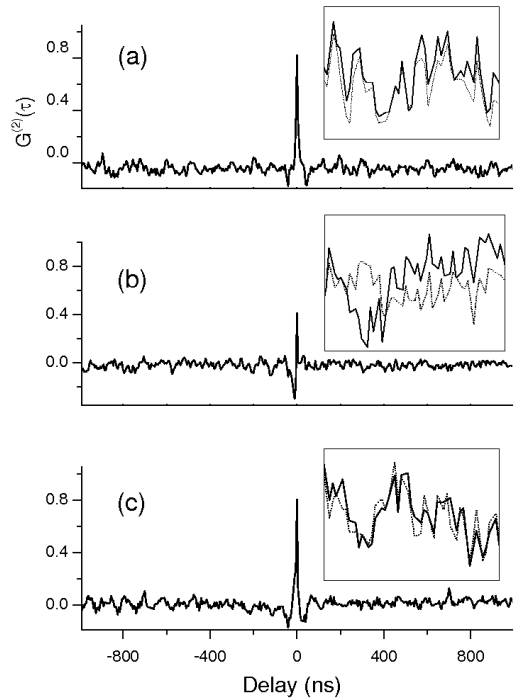}
}
\caption{%Figure 3.
Intensity correlation function $G^{(2)}(\tau)$:  curve (a) recorded
at zero magnetic field, curve (b) at a magnetic field of 0.18 G and
curve (c) at a magnetic field of 0.9 G. Also signals from
photo-detectors are shown in boxes. The waveforms are recorded in
time intervals of 200 ns and amplitude intervals of 4 mW. }
\end{figure}

The signals from the photo-detectors in time intervals of 200 ns
are shown in the inset boxes in Fig.~3a,b,c.
The signals are proportional to the laser beam power with a slope of
500 V/W.
The vertical range of the plot is 4 mV. The amplitude
variations of the signals are practically the same as in Fig.~3.

The temporal behavior of the signal is modified by the applied magnetic field.
The signals shown in Fig.~3a and 3c are correlated, and
the ones shown in Fig.~3b (box) are not.
We observe that
at low and high magnetic field, fluctuations are strongly correlated.
At an intermediate magnetic fields, the peaks of correlation function have
less magnitude. The correlation function behaves as if it changes sign around
zero time delay.
Possibly different frequency components of the signals
can be correlated or anti-correlated at these magnetic fields.
For the dependence shown in Fig.~2c, we have selected the magnitudes
of the correlation function at zero time delay.

We note that the inverse width of the peaks is
the order of $2\pi\cdot 20$~MHz, and it is comparable with the width of
the saturation resonance~\cite{Pappas, Akulshin, chen06jmo},
and it is much broader than the ground state relaxation rate.
The inverse width depends on the excited state
decay rate $2\pi\cdot 6$~MHz and the optical excitation rate.
The influence of one-photon optical saturation on MNOR is discussed in
\cite{BudkerRev2002}.

\section{Theoretical approach and Discussion}

The following is a simplified theoretical description of our system
treating the laser fields classically
and using a density matrix for the atomic response.
The laser beams are in resonance with a three-level medium
as depicted in Fig.~4.

\begin{figure}[tb] %fig2
\center{
\includegraphics[width=.7\columnwidth]{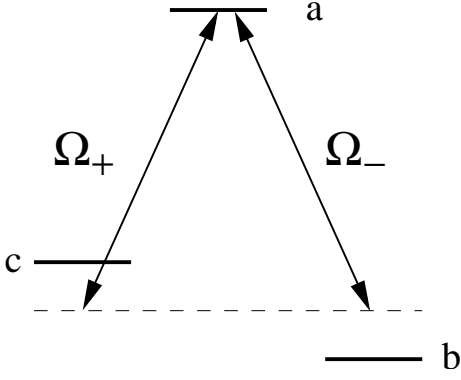}
}
\caption{%Figure 3.
A simplified three-level scheme. Splitting of the ground state is
$\w_{cb} = 2g\mu_B B$, where $g$ is the Lande factor, $\mu_B$ is
Borh's magneton, and $B$ is the magnetic field. }
\end{figure}

The Hamiltonian of the atom is given by
\be
\hat H = \hbar\W_-|a\ket\bra b| + \hbar\W_+|a\ket\bra c| + h.c.,
\ee
where $\W_{\pm}=\wp_{c,b}E_{\pm}/\hbar$ are the Rabi frequencies
of left- and right-circularly polarized beams; $\wp_{c,b}$ and $E_\pm$ are
the corresponding dipole moments of the atomic transitions and
the electric fields. The density matrix equation is given by
\be
\dot \rho = -{i\over\hbar}[H,\rho] -{1\over2}(\hat\G\rho + \rho\hat \G),
\label{[Hr]}
\ee
where $\hat\G$ is the relaxation matrix, introduced to describe relaxation
processes in the atomic medium~\cite{book}.
The equations for field propagation are
\be
\frac{\partial\W_-}{\partial{z}}=-i\eta_b\rho_{ab},
\;\;\;
\frac{\partial\W_+}{\partial{z}}=-i\eta_c\rho_{ac},
\ee
where $\eta_b=\nu_- N \wp^2_{b}/(2\hbar\epsilon_0 c)$,
$\eta_c=\nu_+ N \wp^2_{c}/(2\hbar\epsilon_0 c)$ are the coupling constants,
$\nu_{\pm}$ are the frequencies of circular polarized fields,
$\wp_{b,c}$ are the dipole moments of the corresponding transitions,
$N$ is the density of medium. Note that the left- and right-polarized beams
are the normal modes of the current system, and
their polarizations do not change while they are propagating through the cell.

The intensities of the optical beams propagating through the cell
fluctuate.
The corresponding correlation function is defined as
\be
G^{(2)}(\tau) = {\bra\delta I_-(t) \delta I_+(t+\tau)\ket
\over\sqrt{
\bra [\delta I_-(t)]^2\ket\bra[\delta I_+(t+\tau)]^2\ket
}
}.
%= \int d\delta\w F(\delta\w) I_1(\delta\w) I_2(\delta\w)
\ee
Intensity fluctuations have been experimentally
and theoretically studied in \cite{VAS2005pra, Ari2006}, and
they are related to the atomic responses
\be
\delta I_- \sim \Im \rho_{ab}\W_-^*, \;\;\;
\delta I_+ \sim \Im \rho_{ac}\W_+^*.
\ee
The corresponding atomic coherences in the three-level system
can be found from the solution of the density matrix equations.
Assuming the phase diffusion of optical fields is a slow process,
the atomic coherences are given by
\be
\rho_{ab} = -i{n_{ba}\W_- + \rho_{cb}\W_+\over\G_{ab}},
\;\;\;
\rho_{ca} =  i{n_{ca}\W_+ + \rho_{cb}\W_-\over\G_{ca}}
\label{rab}
\ee
\be
\G_{cb}\rho_{cb} = i \rho_{ca}\W_-  - i \rho_{ab}\W_+,
\label{rcb}
\ee
where
$\G_{ab} = \g_{ab} + i(\w_{ab}-\nu_-)$;
$\G_{ca} = \g_{ca} - i(\w_{ac}-\nu_+)$;
$\G_{cb} = \g_{cb} + i(\w_{cb}-\nu_- + \nu_+)$;
$n_\alpha=\rho_{\alpha\alpha}$;
$n_{\alpha\beta} = \rho_{\alpha\alpha} - \rho_{\beta\beta}$;
$\w_{\alpha\beta}$ are the atomic frequencies; $\alpha$ and $\beta$ are labels
for atomic levels $a,b,c$;
$\nu_{-}(t) = \nu_+(t)$ are the instantaneous frequencies of laser radiation
in both beams having orthogonal polarizations.
Assuming $\W_-=\W_+=\W$, and defining $\delta = \w_{ab} - \nu_- -\w_{cb}/2$,
we solve Eq.~(\ref{rcb}) with respect to $\rho_{cb}$.
Then, substituting the solution into Eqs.~(\ref{rab}) gives us the following:
\be
\rho_{ab} = i{
\G_{cb}(\G -i\delta)n_{ab} + n_{cb}\W^2
\over
\G_{cb}(\delta^2 + \G^2) + 2\G\W^2}\W,
\label{ab}
\ee
\be
\rho_{ca} = i{
\G_{cb}(\G +i\delta)n_{ca} + n_{cb}\W^2
\over
\G_{cb}(\delta^2 + \G^2) + 2\G\W^2}\W.
\label{ca}
\ee
The set of equations for populations can be obtained by substituting
Eqs.(\ref{ab},\ref{ca}) into Eq.(\ref{[Hr]}) for
the appropriate atomic populations:
\begin{widetext}
\be
\g_a n_a + \left(
\ds{\G_{cb}(\G - i\delta)\over\tilde\G_{cb}} +
\ds{\G_{bc}(\G^* + i\delta)\over\tilde\G_{cb}^*}
\right)\W^2 n_{ab}
+
\left(
\ds{1\over\tilde\G_{cb}} + \ds{1\over\tilde\G_{cb}^*}
\right) \W^4 n_{cb} = 0,
\ee
\be
\g_a n_a + \left(
\ds{\G_{cb}(\G + i\delta)\over\tilde\G_{cb}} +
\ds{\G_{bc}(\G^* - i\delta)\over\tilde\G_{cb}^*}
\right)\W^2 n_{ac}
+
\left(
\ds{1\over\tilde\G_{cb}} + \ds{1\over\tilde\G_{cb}^*}
\right) \W^4 n_{bc} = 0,
\ee
\end{widetext}
where $\tilde\G_{cb} = \G_{cb}(\delta^2 + \G^2) + 2\G\W^2$.
By introducing $A_b = A - \delta B$, $A_c = A + \delta B$, and
\be
A = \left(
\ds{\G_{cb}\G\over\tilde\G_{cb}} +
\ds{\G_{bc}\G^*\over\tilde\G_{cb}^*}
\right)\W^2, \;\;\;
B = -i\left(
\ds{\G_{cb}\over\tilde\G_{cb}} -
\ds{\G_{bc}\over\tilde\G_{cb}^*}
\right)\W^2,
\label{AB}
\ee
\be
C = \left(
\ds{1\over\tilde\G_{cb}} + \ds{1\over\tilde\G_{cb}^*}
\right) \W^4,
\label{C}
\ee
the set equations for the populations can be re-written in the compact form
\be
\g_a n_a + A_b n_{ab} + C n_{cb} = 0,
\label{na}
\ee
\be
\g_a n_a + A_c n_{ac} + C n_{bc} = 0,
\label{nc}
\ee
and solved (see Appendix~\ref{app}).
The solution for the population difference
in level $b$ and $c$ is given by
\be
n_{cb} = {2\g_a B\delta
\over
3A_bA_c + C(3A_b+3A_c+4\g_a) + \g_a(A_b+A_c)}.
\label{ncb}
\ee
Simplifying Eqs.~(\ref{AB},\ref{C}), we obtain $A \simeq \g_{cb}$,
$B \simeq {\Delta\over\g}$,  $C \simeq {|\W|^2\over\g}$, and
\be
n_{cb} \simeq {\Delta\delta\over|\W|^2}.
\ee
The intensity fluctuations are determined by atomic coherences, which are
\be
\Im\rho_{ab} \simeq {\g_{cb}\g + \Delta\delta\over\g|\W|}, \;\;\;
\Im\rho_{ac} \simeq {\g_{cb}\g - \Delta\delta\over\g|\W|},
\label{rab-rac}
\ee
where the field phase fluctuation is related to
the frequency deviation $\delta$
that is restricted by the EIT window~\cite{VAS2005pra},
$\delta \sim \ds{\W^2\over\g}$.

We can clearly see from Eq.(\ref{rab-rac})
that the intensity fluctuations have two contributions:
the first originates from absorption, which is the same for both modes,
and the second originates from the Raman term appearing from the population
difference $n_{cb}$, which has opposite signs for these two modes. One field
is amplified due to absorption of the second field.
Depending on the one- and two-photon
detuning, $\Delta$, the intensity fluctuations occur in phase
if $\g_{cb}\g^2/\W^2 \gg \Delta$
or out of phase if $\g_{cb}\g^2/\W^2 \ll \Delta$.
The last condition gives rise to the anti-correlations.

Let us note here that at the EIT condition, $\Delta=0$,
correlated intensities can be also
considered from the point of view of matched pulses~\cite{harris-match}.
The details of a study of switching between
correlations and anti-correlations can be found in \cite{VAS2005pra, Ari2006}

Now we are ready to consider intensity fluctuations
between the orthogonally linearly polarized beams in the current experiment.
Linearly polarized light can be regarded as a linear combination of
left- and right-circular polarized light. The light with left- and
right-circular polarizations do not change their state of polarization
while propagating through the gas cell.
We previously studied the correlation of intention fluctuations
between the normal modes~\cite{VAS2005pra}.

In the current experiment,
to detect rotation of polarization, we use the following scheme.
After the polarizer, the beam propagates through the cell, and then after
a half-wave plate, the polarization plane rotates 45$^o$ degrees
and is split by a PBS into two orthogonally polarized optical beams.
The intensity of each beam is detected.
Thus, the measured signals, $S_1$ and $S_2$,
are proportional to the intensities of the propagated circularly polarized
components, which are given by
\be
S_{1,2}\sim I_{1,2} = {1\over2}(I_+ + I_- \pm 2 \sqrt{I_- I_+} \sin\phi),
\label{s12}
\ee
and the polarization rotation is given by
\be
\phi = \arcsin\left({I_1 - I_2\over I_1 + I_2}\right).
\ee

Then correlation between $S_1$ and $S_2$ can be calculated by
%\begin{widetext}
\beq
G^{(2)}(\tau) = {\bra \delta I_1(t) \delta I_2(t + \tau)\ket
\over
\sqrt{\bra [\delta I_1(t)]^2\ket\bra[\delta I_2(t + \tau)]^2\ket}}.
\label{s1s2}
\eeq
%\end{widetext}

As is clearly seen from Eq.(\ref{s12}) and Eq.(\ref{s1s2}),
the circularly polarized beams are equally split into
two linear polarizations. Thus, the two intensities of the linearly polarized
components are correlated. The only contribution that decreases
the correlations is due to the terms depending on the rotational angle,
which have different signs for orthogonal polarizations in Eq.(\ref{s12}).

The intensity fluctuations should be small for the components that are not
perfectly anti-correlated. But as we have shown in a previous
paper \cite{VAS2005pra}, increasing the magnetic field causes
these components to become
anti-correlated. On another hand, increasing the magnetic field increases
the rotation at first, but then once the splitting becomes bigger than the
EIT width, the rotation angle decreases, restoring the correlations between
the intensities of the linear components.

One can see from Eq.(\ref{s1s2})
that if the rotational angle in a magnetic field
is small, the beams mainly consist of the sum of intensities for left- and
right-circularly polarized beams, and thus the fluctuations are correlated.
Anti-correlations
come from the terms having different signs due to Faraday rotation in
the magnetic field. So, at zero magnetic field, we have only correlations.
Note here that the sum of intensities is a major contribution and always
contributes to correlations between intensities.
But for stronger magnetic fields, due to the decreasing of the angle of
rotation, this term vanishes, and only correlation between beams is observed.

In the experiment, on the other hand, we observe the decreasing of correlation
with increasing magnetic field intensity. It looks similar
to the results obtained for circular polarization, but the physics is
completely different. Actually, the strong decrease of correlations
observed in the experiment occurs because of strong anti-correlations
for higher two-photon detunings.

For larger two-photon detuning, the nonlinear magneto-optical
rotation of polarization
becomes smaller and intensity correlations restore their correlations.
Define
\be
I_+ = I_0 + i_+, \;\;\;
I_- = I_0 + i_-,
\ee
where $i_\pm$ are the intensity fluctuations of the beams $I_\pm$,
correspondingly.
Then, using $\sqrt{1 + z} \simeq 1 + {z\over2}-{z^2\over8}$,
introducing $x = i_+ + i_-$ and
$s=i_+-i_-$ (note that
$i_+i_- - {x^2\over 4} = 4i_+i_- - (i_+ + i_-)^2 = -(i_+ - i_-)^2 = -s^2$),
we can rewrite Eqs.~(\ref{s12}) as
\be
I_{1,2} = 2I_0 + x \pm (2I_o + x - {s^2\over4I_0})\sin\phi.
\ee
Thus we obtain
\be
\delta I_{1,2}=I_{1,2}-\bra I_{1,2}\ket = x(1 \pm \sin\phi) \mp
{s^2 - \bra s^2\ket\over4  I_0}\sin\phi
\ee
and
\be
\bra(I_1-\bra I_1\ket)^2\ket =
\bra x^2\ket(1+\sin\phi)^2 +
{\bra s^4\ket - \bra s^2\ket^2\over16  I_0^2}\sin^2\phi
\ee
Finally, we can calculate correlation function $G^{(2)}$,
defined by Eq.(\ref{s1s2}),
\begin{widetext}
\be
G^{(2)}(\tau) = {\bra x^2\ket\cos^2\phi + \ds{\bra s^4\ket -
\bra s^2\ket^2\over16 I_0^2}\sin^2\phi
\over
\sqrt{
\left(\bra x^2\ket\cos^2\phi +
\left(\ds{\bra s^4\ket - \bra s^2\ket^2\over16  I_0^2}\right)\sin^2\phi
\right)^2 +
4\bra x^2\ket\ds{\bra s^4\ket - \bra s^2\ket^2\over 16  I_0^2}
\sin^4\phi
}}
\ee
\end{widetext}

In the last equation, one can see that
for small magnetic fields, when the rotation angle is small,
and the beam intensities are correlated~\cite{VAS2005pra},
and for strong magnetic fields, when the rotation angle is also small
and the beam intensities are anti-correlated~\cite{VAS2005pra},
the correlation function equals to unity.
These results are independent of correlations or anti-correlations
between the beams. Correspondingly,
in the intermediate case, we have correlations that are less perfect.

The foregoing discussion furnishes
a qualitative description of the correlation behavior
we have observed in this experiment. To obtain quantitative agreement,
we would need
to take into account the intensity fluctuations more accurately, beyond
the small variations as we have assumed here.
We will present the results of such an analysis elsewhere.

%Also let us emphasize the difference of the results reported
%here, in this paper,
%from the results presented in the previous paper \cite{VAS2005pra}.

\section{Conclusions}

We have studied atomic noise correlations in a nonlinear magneto-optical
rotation experiment with Rubidium atomic vapor by using broadband detection.
The correlations between the orthogonally polarized components of the laser
beam are maximal in the absence of a magnetic field. The width of the
correlation function peak is
proportional to the excited state lifetime and the inverse Rabi
frequency. When a longitudinal magnetic field is applied,
the correlations first decrease and then increase.
The minimal correlations
and the maximal rotation angles are observed at the same magnetic fields.
These results can be useful for improving optical magnetometers and
for squeezing optical fields or atomic spins.

We thank M.M. Kash, D.X. Ma, I. Novikova, D. Sarkisyan, A.S.Zibrov,
M.S. Zubairy, and G.R. Welch for useful and fruitful discussions,
H.Chen for his help in
experiment and gratefully acknowledge the support from the Office of Naval
Research, the Robert A.\ Welch Foundation (Grant \#A1261). One of us
(T.S.V) also thanks NFSAT\ award TFP 2005/02 for financial
support and IQS and Department of Physics of Texas A\&M University
for hospitality.

\newpage

\appendix
\section{Solution of set of the density matrix equations}
\label{app}

The set of equations from populations Eq.(\ref{na},\ref{nc}) is given by
\be
\g_a n_a + A_b n_{ab} + C n_{cb} = 0
\label{app-na}
\ee
\be
\g_a n_a + A_c n_{ac} + C n_{bc} = 0
\label{app-nc}
\ee
where $A_b$, $A_c$, $C$ are defined above
(see Eqs.(\ref{AB},\ref{C})).
Then solution of Eqs.(\ref{app-na},\ref{app-nc}) are the following
\be
n_a = {A_bA_c + C(A_b+A_c)
\over
3A_bA_c + C(3A_b+3A_c+4\g_a) + \g_a(A_b+A_c)}
\ee
\be
n_b = {A_bA_c + C(A_b+A_c+2\g_a) + A_c\g_a
\over
3A_bA_c + C(3A_b+3A_c+4\g_a) + \g_a(A_b+A_c)}
\ee
\be
n_c = {A_bA_c + C(A_b+A_c+2\g_a) + A_b\g_a
\over
3A_bA_c + C(3A_b+3A_c+4\g_a) + \g_a(A_b+A_c)}
\ee
and population differences can be presented in the following form
\be
n_{ab} = -{\g_a(A_c + 2C)
\over
3A_bA_c + C(3A_b+3A_c+4\g_a) + \g_a(A_b+A_c)},
\ee
\be
n_{ca} = {\g_a(A_b + 2C)
\over
3A_bA_c + C(3A_b+3A_c+4\g_a) + \g_a(A_b+A_c)},
\ee
\be
n_{cb} = {\g_a(A_b-A_c)
\over
3A_bA_c + C(3A_b+3A_c+4\g_a) + \g_a(A_b+A_c)}.
\ee


\begin{thebibliography}{99}

%%%%%%%%%%%%%%%%%%%%%%%%%%%%%%%%%%%%%%%

\bibitem{Harris} S.E. Harris, Physics Today 50, 36 (1997).

\bibitem{MarangosRev2005} M. Fleischhauer, A. Imamoglu, J.P. Marangos,
%Electromagnetically induced transparency: Optics in coherent media 
Rev. Mod. Phys, 77, 633-673 (2005).

\bibitem{kochar-eit} O. Kocharovskaya, Ya.I. Khanin,
Sov. Phys. JETP 63, 945 (1986).

\bibitem{vas04pra} V.A. Sautenkov, Y.V. Rostovtsev, C. Y. Ye,
G.R. Welch, O. Kocharovskaya,
and M.O. Scully, Phys. Rev. A 71, 063804 (2005).

\bibitem{BudkerRev2002} D. Budker, W. Gawlik, D.F. Kimball, S.M. Rochester,
V.V. Yashchuk, and A. Weis, Rev. Mod. Phys. \textbf{74}, 1153 (2002);
D. Budker, M. Romalis, Nature Phys. 3, 227 (2007);
A. Edelstein, J. Phys.-Cond. Mat. 19, 165217 (2007).

\bibitem{harris-max} M.  Jain, H. Xia, G.Y. Yin, et al.,
Phys. Rev. Lett. 77, 4326 (1996).

\bibitem{vas04pra2} V. A. Sautenkov, C. Y. Ye, Y. V. Rostovtsev,
G. R. Welch, and M. O. Scully,
Phys. Rev. A 70, 033406 (2004).

\bibitem{Weis1992}
A. Weis, V.A. Sautenkov and T.W. Hansch,
Phys. Rev.A 45, 7991 (1992).

%\bibitem{scully99prl} M.O. Scully, Phys. Rev. Lett. 67, 1855 (1991).

\bibitem{Hau1999} L.V. Hau, S.E. Harris, Z. Dutton, C.H. Behroozi,
Nature, 397, 594 (1999).

\bibitem{Kash1999} M.M. Kash, V.A.Sautenkov, A.S. Zibrov, et al.
Phys. Rev. Lett. 82, 5229,  (1999).

\bibitem{Budker1999} D. Budker, D.F. Kimball, S.M. Rochester, et al.,
%Nonlinear magneto-optics and reduced group velocity of light in atomic vapor with slow ground state relaxation 
Phys. Rev. Lett. 83, 1767, (1999).

\bibitem{delay} E.E. Mikhailov, V.A. Sautenkov, Y.V. Rostovtsev,
and G.R. Welch, J. Opt. Soc. Am. B 21, 425 (2004);
Q. Sun, Y.V. Rostovtsev, J.P. Dowling, M.O. Scully, and
M. S. Zubairy, Phys. Rev. A 72, 031802 (2005).

\bibitem{sound} A. B. Matsko, Y. V. Rostovtsev, M. Fleischhauer,
and M. O. Scully, Phys. Rev. Lett. 86, 2006 (2001).

\bibitem{rost06prl} Y.V. Rostovtsev, Z.-E. Sariyanni, and M.O. Scully,
Phys. Rev. Lett. 97, 113001 (2006).


\bibitem{switching} C. Y. Ye, V. A. Sautenkov, Y. V. Rostovtsev,
and M. O. Scully, Opt. Lett. 28, 2213 (2003).

\bibitem{kalugin06ol} N.G. Kalugin, Y. Rostovtsev,
Opt. Let. 31, 969 (2006); E. Kuznetsova, Y. Rostovtsev, N.G. Kalugin,
R. Kolesov, O. Kocharovskaya, and Marlan O. Scully,
Phys. Rev. A 74, 023819 (2006).

\bibitem{Hau2001}
%Observation of coherent optical information storage in an atomic medium using halted light pulses,
C. Liu, Z. Dutton, C.H.  Behroozi, L.V. Hau,
Nature, 409, 490  (2001).

\bibitem {Lukin2001} D.F.Phillips, A.Fleischhauer, A.Mair et al,
Phys. Rev. Lett. 86, 783 (2001).


\bibitem{Harris1998} S.E. Harris, Y. Yamamoto,
Phys. Rev. Lett. 81, 3611 (1998).


\bibitem{lukin-hemmer} M.D. Lukin, P.R. Hemmer, M.O. Scully,
Adv. in At. Mol. Opt. Phys. 42, 347 (2000).



\bibitem{scully01prl} M.O. Scully, M. Fleischhauer,
Phys. Rev. Lett. 69, 1360 (1992);
M. Fleischhauer, M.O. Scully,
Phys. Rev. A49, 1973 (1994);
V.A. Sautenkov, M.D. Lukin, C.J. Bednar CJ, et al.,
%Enhancement of magneto-optic effects via large atomic coherence in optically dense media
Phys. Rev. A 62, 023810 (2000);
M. Fleischhauer, A.B. Matsko, M.O. Scully,
%Quantum limit of optical magnetometry in the presence of ac Stark shifts
Phys. Rev. A 62, 013808 (2000).


\bibitem{Budker2001pra}
S. Rochester, D.S. Hsuing, D. Budker, R.Y. Chiao, D. Kimball, V. Yashchuk,
%Self-rotation…
Phys. Rev. A 62, 043403 (2000).

\bibitem{Weis2005ELO} A. Weis, R. Wynands,
%Laser-based precision magnetometry in fundamental and applied research, 
%OPTICS AND LASERS IN ENGINEERING
Opt. Las. in Eng., 43,  387,  (2005).

\bibitem {Hollberg2005} P.D.D. Schwindt, L. Hollberg, J. Kitching,
%Self-oscillating rubidium magnetometer using nonlinear
%magneto-optical rotation,
Review of Scientific Instruments, 76, 126103 (2005).

\bibitem{Budker2000pra} D. Budker, D. Kimball, S. Rochester,
V. Yashchuk and M. Zolotarev,
%Sensitive magnetometry based on nonlinear magneto-optical rotation
Phys. Rev. A 62, 043403 (2000).

\bibitem{VAS2000}
V. A. Sautenkov, M. D. Lukin, C. J. Bednar, I. Novikova, E. Mikhailov, M.
Fleischhauer, V. L. Velichansky, G. R. Welch, and M. O. Scully,
Phys. Rev. A 62, 023810 (2000).


\bibitem{Yabuzaki1991} T. Yabuzaki, T. Mitsui, and U. Tanaka,
%NEW TYPE OF HIGH-RESOLUTION SPECTROSCOPY WITH A DIODE-LASER
Phys. Rev. Lett. 67, 2453 (1991).

\bibitem{Camparo1998} J.C. Camparo,
%Conversion of laser phase noise to amplitude noise in an optically thick vapor.
%\emph{It was shown that conversion of phase noise to intensity noise drops with
%laser intensity.}
J. Opt. Soc. Am. B 15, 1177 (1998).

\bibitem{Camparo1999} J.C. Camparo and J.G. Coffer,
%Conversion of laser phase noise to amplitude noise in a resonant atomic vapor:
%The role of laser linewidth
Phys. Rev. A 59, 728 (1999).

\bibitem{Walser1993} D. H. McIntyre, C. E. Fairchild,
J. Cooper, and R. Walser,
%Diode-laser noise spectroscopy of rubidium
Opt. Lett. 18, 1816 (1993).

\bibitem{Boyd1995}
%Excess noise acquired by a laser beam after propagating through an
%atomic-potassium vapor/ W. V. Davis, M. Kauranen, E. M. Nagasako,
R. J. Gehr, A. L. Gaeta, and R. W. Boyd, and G.S.Agarwal,
Phys. Rev. A 51, 4152 (1995)

\bibitem{Zibrov2005}
%Observation of quantum noise in the polarization of laser light in a rubidium-vapor cell
A.S. Zibrov, I. Novikova, JETP Let. 82, 110 (2005)

\bibitem{Ari2006} G.O. Ariunbold, V.A. Sautenkov, Y.V. Rostovtsev,
M.O. Scully, arXiv:quant-ph/0603025.

\bibitem{AlzarEPL2003}
%Super-Poissonian intensity fluctuations and correlations between
%pump and probe fields in Electromagnetically Induced Transparency
C.\ L.\ Garrido-Alzar, L.\ Cruz, J.\ Aguirre-Gomez, M.\ F.\ Santos, and P.\  Nussenzveig,
Europhys.\ Lett.\ \textbf{61}, 485 (2003).

\bibitem{Lezama2006}
L. S. Cruz, D. Felinto, J.G.A. Gomez, M. Martinelli, P. Valente, A. Lezama,
and P. Nussenzveig,
Eur.Phys.J. D, 41, 531 (2007).

\bibitem {VAS2005pra} V.A. Sautenkov, Yu. Rostovtsev and M.O. Scully,
%Swithching between photon-photon correlations and Raman anticorrelations in a coherently prepared Rb vapor
Phys. Rev. A 72, 065801 (2005).

\bibitem{vas07jmo} V.A. Sautenkov, H. Li, Yu. Rostovtsev and M.O. Scully,
J. Mod. Opt. 54, 2451 (2007).

\bibitem{Lezama2004}
M. Martinelli, P. Valente, H. Failache, D. Felinto, L.S. Cruz, P. Nussenzveig,
and A. Lezama, Phys. Rev. A 69, 043809 (2004).
%"Noise spectroscopy of nonlinear magneto-optical resonances in Rb vapor"


\bibitem{Roch1997} J.F.Roch, K.Vingeron, Ph.Grelu et al,
Phys. Rev. Lett. 78, 634 (1997).

\bibitem {Lukin1999}
M.D.Lukin, A.B.Matsko, M. Fleischhauer, and M.O.Scully
Phys. Rev. Lett. 82, 1847 (1999).

\bibitem{Pinard2006prl} A. Dantan , J. Cviklinski J, E. Giacobino, et al.
%Spin squeezing and light entanglement in coherent population trapping 
Phys. Rev. Lett. 97, 023605 (2006).

%\bibitem{Matsko2002pra} [Matsko2002pra,

\bibitem{Matsko2002}
%Vacuum squeezing in atomic media via self-rotation
A.B. Matsko, I. Novikova, G.R. Welch, D. Budker, D.F. Kimball,
S.M. Rochester, Phys. Rev. A66, 043815 (2002).

\bibitem{Matsko2003}
%\textbf{Sq in EIT}
%Nonlinear magneto-optical rotation of elliptically polarized light"
A.B. Matsko, I. Novikova, M.S.Zubairy, G.R. Welch,
Phys. Rev. A 67, 043805 (2003)

\bibitem{Lvovsky2003} J. Ries, B. Brezger, A.I. Lvovsky,
%Experimental vacuum squeezing in rubidium vapor via self-rotation,
Phys. Rev. A 68, 025801 (2003).

\bibitem{Pinard2006pra}
M.T.L. Hsu, G. Hetet, A. Peng, et al.,
%Effect of atomic noise on optical squeezing via polarization self-rotation in a thermal vapor cell 
Phys.Rev. A 73, 023806 (2006).

\bibitem{Hsu2006prl} M.T.L. Hsu, G. Hetet, O.Glockl, et al.
%Quantum study of information delay in electromagnetically induced transparency 
Phys. Rev. Lett. 97, 183601 (2006).


\bibitem{vasilliev2006} V.V. Vassiliev, S.A. Zibrov,
V.L.Velichansky, Review of Scientific Instruments, 77, 013102 (2006).

\bibitem{Pappas}
P.G. Pappas, M.M. Burns, D.D. Hinshelwood, M.S. Feld, and D.E. Murnick,
%"Saturation spectroscopy with laser optical pumping in atomic barium,"
Phys. Rev. A 21, 1955 (1980).

\bibitem{Akulshin} A.M. Akulshin, V.A. Sautenkov, V.L. Velichansky, et al.,
%POWER BROADENING OF SATURATION ABSORPTION RESONANCE ON THE D2 LINE OF RUBIDIUM
Opt. Commun. 77, 295 (1990).


%======================

\bibitem{chen06jmo} H. Chen, H.B. Li, Y. Rostovtsev, et al.,
%M.A. Gubin, V.A. Sautenkov, M.O. Scully,
JOSA B 23, 723 (2006).

\bibitem{book} M. O. Scully and M. S. Zubairy,
{\it Quantum Optics} (Cambridge University Press, Cambridge, England, 1997).

\bibitem{harris-match} S. E. Harris, Phys. Rev. Lett. 70, 552 (1993).


\end{thebibliography}
\end{document}